\documentstyle[11pt, aaspp4,psfig, tighten]{article}

\def\amin{\ifmmode^{\prime}\else$^{\prime}$\fi}
\def\asec{\ifmmode^{\prime\prime}\else$^{\prime\prime}$\fi}

\def\simgt{\lower.5ex\hbox{$\; \buildrel > \over \sim \;$}}
\def\simlt{\lower.5ex\hbox{$\; \buildrel < \over \sim \;$}}

\newcommand\rosat{{\it ROSAT}}
\newcommand\hst{{\it HST}}
\newcommand\asca{{\it ASCA}}
\newcommand\as{$^{\prime\prime}$}
\newcommand\ha{H$\alpha$}
\newcommand\hii{H{\small II}}

\begin{document}

\title{\large Structure and Evolution of Hot Gas in 30 Dor}
\author{Q. Daniel Wang}
\affil{Dept. of Physics \& Astronomy, Northwestern University}
\affil{2145 Sheridan Road, Evanston,~IL 60208-3112}
\affil{Electronic mail: wqd@nwu.edu}

\begin{abstract}
We have investigated the structure and evolution of hot gas in the 30 Dor 
nebula, based on recent X-ray observations. Our deep \rosat\ HRI 
image shows that diffuse X-ray emission arises in blister-shaped regions 
outlined by loops of \hii\ gas. X-ray spectroscopic data from \asca\ confirm 
the thermal nature of the emission and indicate that hot gas temperature 
decreases from the core to the halo of the nebula. The structure of the nebula 
can be understood as outflows of hot and \hii\ gases from the parent giant 
molecular cloud of the central OB association. The dynamic mixing between the 
two gas phases is likely responsible for the mass loading to the hot gas, as
required to explain the observed thermal structure and X-ray luminosity of 
the nebula. Such processes should also be important in the formation of 
similar giant \hii\ regions and in their subsequent evolution into supergiant
bubbles or galactic chimneys.  

\end{abstract}


\section{Introduction}

	As the most luminous \hii\ nebula in the Local Group of galaxies,
30 Dor in the Large Magellanic Cloud (LMC) provides the most accessible 
paradigm to understand massive star formation and its interaction with the 
interstellar medium (ISM). Responsible for the ionization  
of this spectacular nebula ($\sim 10^{52} {\rm~Ly~photon~s^{-1}}$) 
is the OB association NGC 2070. The central cluster 
R136 ($r \lesssim 10$~pc; assuming the LMC distance as D=47~kpc; Gould 1995) 
alone accounts for about half of the required ionization radiation  
(Walborn \& Blades 1997; Crowther \& Dessart 1998 
and references therein). The bulk of very massive stars,
contained primarily within the compact core of the cluster ($r \sim 2$~pc),
were formed in a star forming burst about 1-2 Myrs ago (Massey \& Hunter 1998).
\hii\ filaments form shells or loops that often extend
more than 100~pc away from the central cluster (e.g., Chu \& Kennicutt 
1994). However, no model has yet been proposed to explain the
origin and evolution of this nebula structure.

	The 30 Dor nebula is the first giant \hii\ region (GHR)
shown to be a strong diffuse X-ray emitter. As seen in an IPC image of the 
{\sl Einstein} Observatory (Wang \& Helfand 1991), the emission peaks within 
\hii\ shells, indicating the presence of hot gas of a 
few times $10^6$~K. Stellar winds from massive stars and supernova explosions 
are presumably responsible for the heating of the gas. Relatively recent 
observations with the {\sl ROSAT} Positional Sensitive Proportional Counter 
(PSPC) have confirmed the {\sl Einstein} results (Norci \& \"Ogelman 1993; 
Chu 1993).

	In this Letter, we present a study of 30 Dor based chiefly on a deep 
X-ray image from the {\sl ROSAT} High Resolution Imager (RHRI) and on an 
X-ray spectrum from the Advanced Satellite for Cosmology and Astrophysics 
(\asca).
These two complimentary data sets enable us to spatially resolve the detailed
X-ray emission structure and to characterize the thermal properties of the 
hot gas. We compare the X-ray data with optical and UV observations and 
extend our discussion (Wang 1996) about the origin of the hot gas and 
its subsequent evolution. 

\section{Description of the X-ray Data}

	 The RHRI image, as presented in Fig. 1, is a co-add of three
observations: \rosat\ \# rh600228n00 
(30~ks exposure),  rh400779a01 (79~ks), and rh400779n00 (26 ks). 
The first observation has already been used in the study of the two bright 
point-like sources in the core of the nebula, which are Wolf-Rayet + black 
hole binary candidates (Fig. 2; R140 and Mrk 34; Wang 1995), and the 
Crab-like supernova remnant N157B (Wang \& Gotthelf 1998a). The second 
observation has been utilized for both timing 
and positioning of the recently discovered 16~ms pulsar within the remnant
(Wang \& Gotthelf 1998b and references therein). Briefly, the RHRI data have 
a spatial resolution ranging from $\sim 6$\as\ (FWHM) near the image center to 
$\sim 30^{\prime\prime}$ near the edges, or a factor of up to $\sim$ 10 
(5) better than the resolution of the IPC (PSPC). The energy coverage is 
in the range of 0.1 - 2~keV. We processed the RHRI data with a software 
provided by S. Snowden
for data editing, background subtraction, and exposure correction. We then
adaptively smoothed the resultant X-ray intensity image with a Gaussian of 
adjustable size to achieve a uniform signal-to-noise ratio of 6 
for Fig. 1 and 4 for Fig. 2.

	The X-ray spectral data were obtained from the \asca\ SIS observation 
\# ad20000000. Wang \& Gotthelf (1998a) have presented two broad 
band images constructed from this observation, which had a spatial
resolution of $\sim 1^\prime$ (FWHM) and a spectral resolution of $\delta E/E 
\sim 0.02$$ (5.9~{\rm keV}/E)^{0.5}$. The observation, taken in a 4-CCD mode,
was pointed at R.A., Dec. (J2000)$ = 5^h38^m32^s, $$
-69^\circ9^\prime44^{\prime\prime}$ and had a roll angle of $20^\circ.3$ north 
to east.  We utilized only the data from the two northern CCD chips of each
sensor to minimize the contamination caused by scattered counts from N157B, 
PSR B0540$-$69, and LMC X-1 to the south. Specifically, the on-nebula spectrum
was extracted from a half circle of 7\farcm5 radius, centered at the pointing 
direction, plus a northern stretch of 5\farcm4 wide to complete the coverage 
of the northern X-ray spur of the nebula (Fig. 1). We estimated the 
background from the same on-nebula chip areas in the adjacent observation 
(ad90001000 center at $5^h43^m31^s, $$
-69^\circ13^\prime48^{\prime\prime}$), just east of 
30 Dor. We also tested an alternative background estimate from the regions 
 away from the nebula, but in the same on-nebula CCD chips. The resultant 
background-subtracted spectra are essentially the same; the differences
in best-fit spectral parameters (Table 1) are all within 5\%. We added the 
spectra from both sensors of each observation (on- or off-nebula)
for subsequent spectral 
model fitting. This spectral data reduction followed the procedure 
prescribed in the on-line \asca\ Data Reduction Guide
(http://heasarc.gsfc.nasa.gov/docs/asca/abc/abc.html).

\section{Results}

The X-ray emission from 30 Dor is predominantly diffuse in origin.
Fig. 1 shows no evidence for point-like X-ray sources which may be 
related to 30 Dor, except for the two X-ray binary candidates mentioned above
(\S 2) and the apparent X-ray contribution from R136 (Fig. 2). In general, 
there is no detailed correlation between the X-ray emission
and the presence of massive stars, as appear in the Astro-1 UIT UV image
(Fig. 3; Hill et al. 1993). The detection limit for an individual point-like
source is $\sim 1 \times 10^{35} {\rm~ergs~s^{-1}}$ (0.5-2~keV). 
Enhanced diffuse X-ray emission is enclosed within
\hii\ loops, which are typically anchored to dense molecular 
clouds in the core of the nebula (Wang 1996; Fig. 3). But the detailed 
morphology of individual features remains uncertain. It is still difficult
to disentangle various projection effects and differential X-ray absorption 
across the field. For example, the two shell-like 
features of $\sim 3^\prime$ diameters 
to the south of R136 may well be the projection of two loops that originate
in the core of the nebula. Apparently, the 30 Dor nebula is a complex of 
diffuse X-ray-emitting blisters.

	In the central region (Fig. 2), strong X-ray emission arises
in cavities outlined by ionization fronts (Scowen et al. 1998) at boundaries 
of molecular clouds (e.g., Wang 1996; Johansson et al. 1998;
Rubio et al. 1998). 
The cavity around R136, in particular, is a well-defined blister,
which is open, at least, to the east (e.g., Dickel et al. 1994). 
\hii\ gas kinematics (e.g, Chu \& Kennicutt 1994) show that the eastern part 
of this blister is expanding much more rapidly 
than the west part. This expansion is most likely driven by 
X-ray-emitting gas that is flowing out from near R136 to the halo of the 
nebula. The transition zone between the core and the halo of the nebula
appears at $r \sim 2^\prime-4^\prime$ from R136, at outer boundaries of the 
parent giant molecular cloud (GMC) of the OB association. Particularly in the 
east-west direction, \ha-emitting filaments clearly show the morphology of
outward streamers (Fig. 3).

	 The \asca\ spectral data (Fig. 4) confirm that X-ray emission in the 
0.5-2~keV band is predominantly thermal in origin. Emission lines are
clearly present: e.g., Ne ($\sim 1.0$~keV), Mg (1.3~keV), and Si (1.9~keV).
The flat spectrum at higher energies, however, suggests the presence of 
a nonthermal component, which may represents the expected hard X-ray 
spectral tails of the black hole binaries (Wang 1995). A power law with
a typical photon index of 2 describes this component well, although its exact 
spectral shape is uncertain due to the limited energy coverage and statistics 
of the data. Using the Raymond \& Smith  optically thin 
plasma model to characterize the thermal portion of the \asca\ spectrum, 
we find that at least two temperature components are needed for a reasonable 
good fit to the overall spectral shape.  But, this three-component 
model is still an oversimplification. Even when the abundances are allowed to 
adjust, the model does not adequately account for several line features which
peak at intermediate plasma temperatures of a few times $10^6$~K. The 
inclusion of these lines in the model (Table 1) gives an acceptable fit to the 
data ($\chi^2/n.d.f. = 124/107$; Fig. 4). The two thermal components 
characterize a minimum temperature range of the X-ray-emitting gas,  
$\sim 2-9 \times 10^6$~K. The high temperature component dominates in the 
energy band $\gtrsim 1.5$~keV, and 
arises primarily in the core region around R136 ($r \lesssim 1^\prime$;
Wang \& Gotthelf 1998a). Howvever, this component cannot be explained by the 
central two-point like sources (Fig. 2); their total RHRI count rate is $0.013 
\pm 0.001$ (Wang 1995), or only a quarter of the predicted count rate of the 
component alone. Therefore, the hard temperature component also
originates predominately in diffuse hot gas and its temperature apparently 
decreases from the core to the halo of the nebula. 

	The X-ray spectral analysis further provides a measurement
of the mean X-ray-absorbing gas column density toward 30 Dor. A comparison of
the measured column density  N$_{\rm H}$ (Table 1) with the mean reddening 
E$(B-V) = 0.4-0.5$ of 30 Dor (e.g., Parker 1993; Dickel et al. 1994) gives
N$_{\rm H}$/E$(B-V) \approx 3.1 \times 10^{22}
{\rm~cm^{-2}~mag^{-1}}$, which is considerably greater than
N$_{\rm HI}$/E$(B-V) \approx 2.4 \times 10^{22} {\rm~cm^{-2}~mag^{-1}}$ 
(Fitzpatrick 1986). This is expected because N$_{\rm H}$
includes contributions from both partially-ionized and molecular gas phases,
in addition to the \ion{H}{1} column density N$_{\rm HI}$. 
The X-ray absorption in the \asca\ energy range ($\gtrsim 0.5$~keV)
is insensitive to the ionization and chemical states of the gas 
(Morrison \& McCammon 1983). 

	Using the best-fit spectral model (Table 1), we estimate physical 
parameters of hot gas in the nebula. 
An approximate conversion between 
the observed RHRI intensity to the emission measure of hot gas is 
$\sim 40 {\rm~cm^{-6}~pc/(counts~ks^{-1}~arcmin^{-2}})$ with an 
uncertainty of $\sim 2$ within the ranges of the spectral parameters. The 
total count rate from a circle of 10$^\prime$ radius around
R136, excluding the N157B and point-like source contributions, 
is $\sim 0.26{\rm~counts~s^{-1}}$. This rate corresponds to an intrinsic 
luminosity $L_x(0.5 - 2~{\rm~keV}) \sim 9 \times 
10^{37} {\rm~ergs~s^{-1}}$ and a radiative cooling rate of $\sim 1\times 
10^{39} {\rm~ergs~s^{-1}}$. The spectral parameters of the high and low 
temperature thermal components suggest an average pressure $p/k \sim 2 
\times 10^7 {\rm~K~cm^{-3}}$ in the core and a factor of $\sim 7$ lower 
in the halo of the nebula. The total thermal energy and mass of
the hot gas, if within a sphere of $r=140$~pc and with a volume filling 
factor $f_{0.2}$ (in units of 20\%), are $\sim (3 \times 10^{52} 
{\rm~ergs})~ f_{0.2}^{1/2}$ and $(4 \times 10^4 M_\odot) f^{1/2}_{0.2}$, 
contained mostly in the low temperature component.

\section{Discussions}
 
The formation of the giant 30 Dor nebula is clearly driven by the intense
energy release from massive stars. With a total mass losing rate of 
$\sim 10^{-3} M_\odot 
{\rm~yr^{-1}}$ and a typical terminal wind velocity of $\sim 3 \times 10^3
{\rm~km~s^{-1}}$, the central cluster R136 alone has a stellar wind luminosity 
of $\sim 3 \times 10^{39} {\rm~ergs~s^{-1}}$ (e.g., Chu \& Kennicutt 1994). 
Over its lifetime, 
the cluster should have released $\sim 10^{53}$~ergs of mechanical energy.
A substantial fraction of this energy is 
contained in the hot gas, and that the radiative cooling of the hot gas
is considerable (\S 3). The remaining energy 
could be accounted for by the global and turbulent motion of various gas 
components (e.g., Chu \& Kennicutt 1994) and by the energy loss to the 
formation of dense gas shells or filaments (Mac Low \& McCray 1988). 
In comparison, the hot gas in the central cavity accounts for only 
$\sim 10^{51}$~ergs. Thus, outflows must have occurred. Through a highly 
inhomogeneous and clumpy medium, the outflows can naturally lead to the 
formation of X-ray-emitting blisters of various shapes and kinematics
(Wang 1996). 

	The key process invloved in the evolution of the hot gas
is mass loading. Just behind the terminal shock, 
stellar wind materials have a temperature of $\sim 10^8$~K, which is one order 
of magnitude greater than the measured average temperature of 
hot gas even in the core of the nebula. The mass loading can naturally
increase the density and decrease the temperature of the materials, 
neccessary for explaining the observed X-ray emission.

	The most effective mass loading process is likely the dynamic mixing 
of hot gas with \hii\ gas. While the mass efficiency of 
massive star formation is typically only 5-10\% (e.g.,  McKee 1989), much of 
the parent GMC of the OB association is being eroded via photon-evaporation. 
The erosion has been accellerated greatly since the nebula broke out from 
the GMC. The mass eroding rate can be estimated as (Whitworth 1979; 
York et al. 1989) $\sim 0.12 M_\odot {\rm~yr^{-1}}$$ (S_{i}/10^{52} 
{\rm~Ly~photons~s^{-1}})^{4/7} $$
(t_\ast/10^6 {\rm~yr})^{2/7}$$ (n_o/10^3 {\rm~cm^{-3}})^{-1/7},$
where $S_i$, $t_\ast$, and  $n_o$ are the ionizing flux of NGC 2070, the 
effective age of the OB association with the present flux, and 
 the mean density of the GMC. Only if $\sim 10\%$ of this photon-evaporated
\hii\ gas is loaded to the shocked wind materials in
the core of the nebula, can the temperature of hot gas there be explained.
Over the lifetime of the nebula, the total evaporated mass is 
$\sim 10^5 M_\odot$. In order to account for the total mass of hot gas 
($\sim 4 \times 10^4 M_\odot$; \S 3), the mass loading over the nebula must 
be substantial.

	The dynamic mixing is evident, right within 
the central cavity around R136. \hst\ WFPC2 images of optical 
emission lines clearly demonstrate the presence of evaporative flows from 
various ionization fronts into the interior of the cavity (Scowen et al. 1998).
Such flows are caused by the lack of a
pressure confinement of \hii\ gas after hot gas has largely escaped 
from the GMC through outflows. Both the interaction between 
stellar winds and evaporative flows and the interface between the outflows 
of hot and \hii\ gases are highly unstable. The dynamic mixing can naturally
occur. 

	Supernova explosions may also play a role in the 
formation of the 30 Dor nebula. To be energetically important, however, the 
number of explosions needs to be several tens, at least, if each 
releases typically a few times $10^{50}$~ergs of mechanical energy.
Supernova blastwaves can sporadically shock large amounts of evaporated \hii\ 
gas in the core. The relatively high pressure there tends to 
drive the heated gas out into the halo of the nebula, similar to the outflows
of mass-loaded stellar wind materials. 

	This study of 30 Dor leads us to conclude that outflows of GHRs from 
GMCs can naturally convert large amounts of molecular gas into hot gas,
via photon-evaporation, dynamic mixing, and occasionally shock-heating. 
Such mass loading can greatly enhance
the radiative cooling of the hot gas, therefore the X-ray luminosities of 
the GHRs. The hot gas properties of supergiant bubbles and galactic chimneys
can also be affected, as they are all evolved from GHRs and have most likely 
experienced the outflow phase.

\begin{acknowledgements}
{\noindent \bf Acknowledgments} --- We thank Y.-H. Chu for her comments
on the manuscript. This research was supported partly by NASA LTSA grant 
NAG5-6413. 
\end{acknowledgements}
\vfil
\eject

\begin{deluxetable}{lr}
\small
\tablewidth{0pt}
\tablecaption{Results of Spectral Analysis$^a$} 
\tablehead{Parameters & Values}
\startdata
Photon Index 		& 2 (fixed)\\
\ norm 			& $3.9 \times 10^{-4}$\\
Low temperature 	&0.150(0.145-0.159)~keV	\\
\ norm 			&0.94\\
High temperature 	& 0.81(0.73-0.89)~keV\\
\ norm	 		&$9.7 \times 10^{-3}$\\
Fe L-shell 		&0.71(0.62-0.80)~keV\\
\ Gaussian Width	&$< 9.5\times 10^{-2}$~keV\\
Ne Line 		&1.03(0.99-1.06)~keV\\
\ Gaussian Width 	&$< 3.1\times 10^{-2}$~keV\\
Si Line 		&1.35(1.34-1.36)~keV\\
\ Gaussian Width 	&$< 2.6\times 10^{-2}$~keV\\
Column density 		&$1.43 (1.35-1.48)\times  10^{22} {\rm~cm^{-2}}$\\
\enddata
\tablenotetext{a} {The spectral normalizations are in the standard XSPEC units:
${\rm~photons}$${\rm~keV^{-1}}$${\rm~cm^{-2}~s^{-1}}$ at 1~keV for the 
power law and $[10^{-14}/(4\pi D^2)]$$\int n_e n_H dV$ for the two thermal 
plasma components, where $D$ 
is the distance to 30 Dor, $n_e$ and $n_H$ are the electron and hydrogen 
densities (all in cgs units). The metal abundances are assumed to be 30\% of 
the solar values for both X-ray-emitting and absorbing gases.
The uncertainty intervals are all at the 90\% confidence.}
\end{deluxetable}
\vfil
\eject

\begin{figure}
\caption{RHRI X-ray image of 30 Dor. The contours are at 0.6, 1.2, 1.9, 3.0, 
4.5, 6.4, 9.1, 15, 24, 39, 63, 100, 160, 256, 407, and $647 \times 10^{-3} 
{\rm~counts~s^{-1}~arcmin^{-2}}$, above a local background of 
$\sim 1 \times 10^{-3}$.} 
\end{figure} 

\begin{figure}
\caption{The central region of 30 Dor in X-ray and \ha. 
The RHRI contours start at $ 3 \times 10^{-3} {\rm~counts~s^{-1}~arcmin^{-2}}$.
The rest is the same as Fig. 1.
} 
\end{figure} 

\begin{figure}
\caption{Comparison of 30 Dor in \ha\ (green), UV (blue), and X-ray (red). } 
\end{figure} 

\begin{figure}
\caption{\asca\ SIS spectrum of 30 Dor. Individual spectral components
are plotted separately (Table 1): the power law (pink), the low and high 
temperature thermal components (blue and red), and the three Gaussian 
emission lines (green).}
\end{figure} 
\vfil
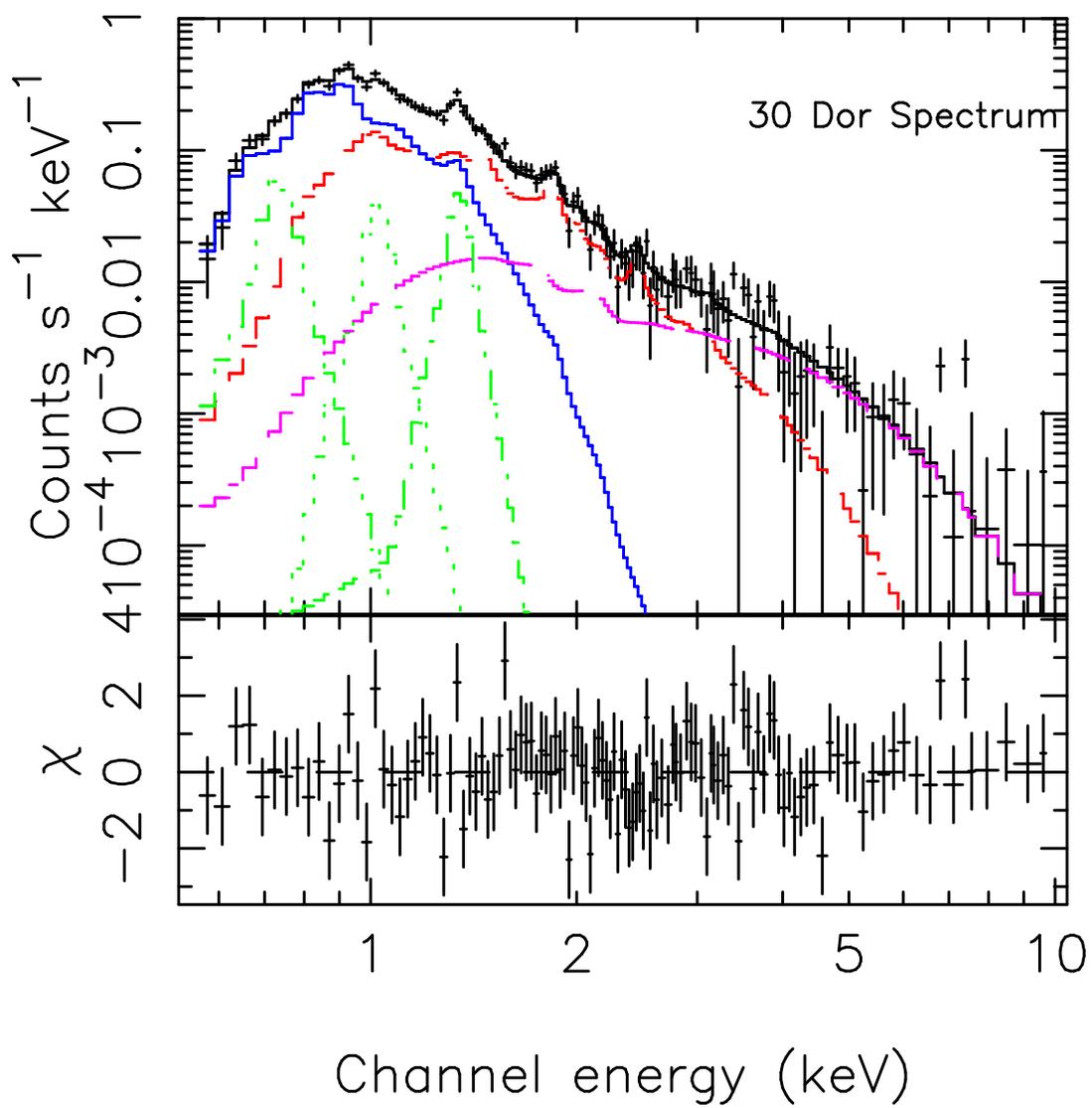
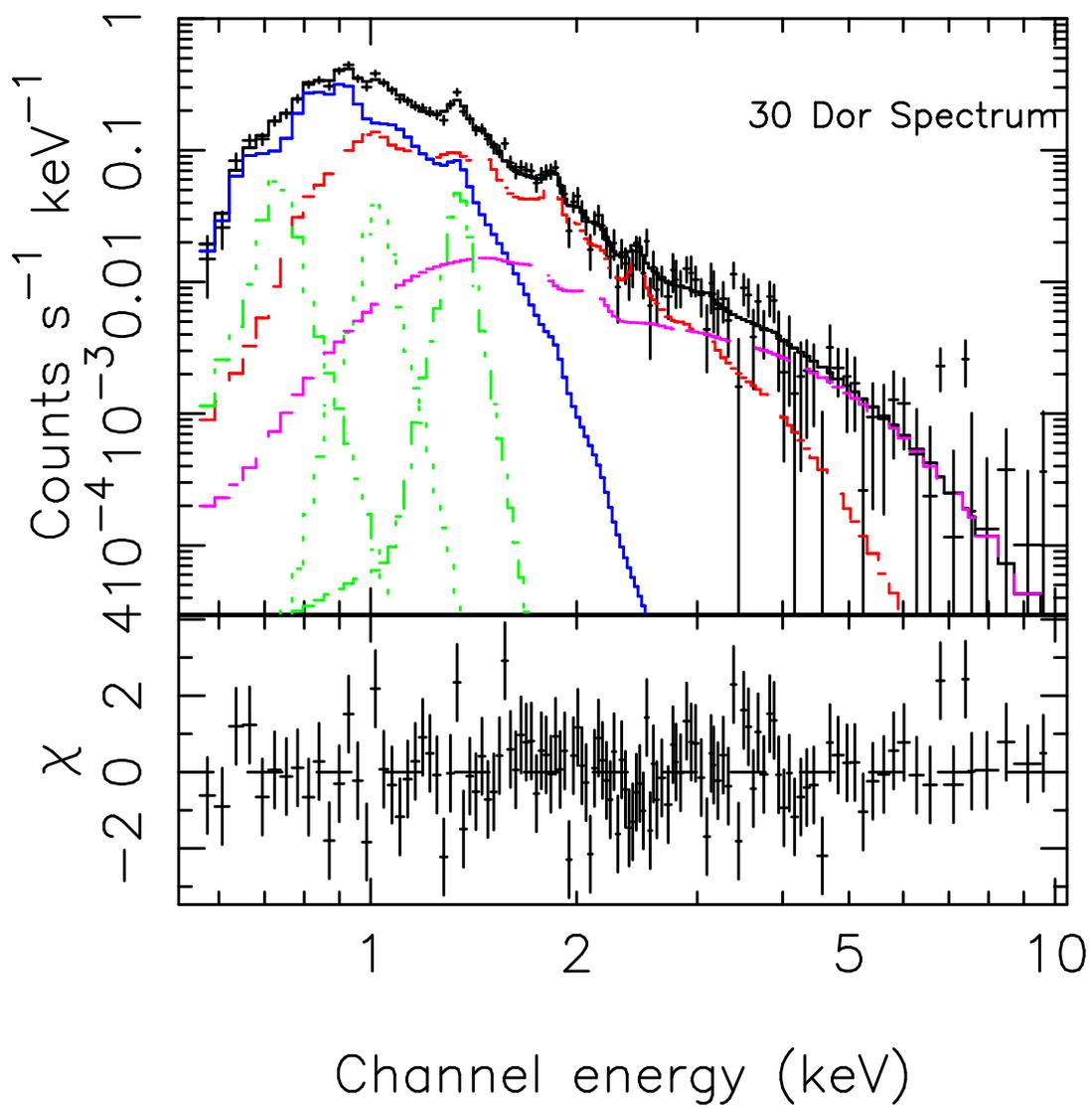
\end{document}